\documentclass[prl,twocolumn,showpacs,preprintnumbers,amsmath,amssymb]{revtex4}

\usepackage{graphicx}
\usepackage{color}
\usepackage{epsfig}
\usepackage{latexsym}
\usepackage{bm}

\begin{document}

\renewcommand{\ni}{{\noindent}}
\newcommand{\dprime}{{\prime\prime}}
\newcommand{\be}{\begin{equation}}
\newcommand{\ee}{\end{equation}}
\newcommand{\bea}{\begin{eqnarray}}
\newcommand{\eea}{\end{eqnarray}}
\newcommand{\nn}{\nonumber}
\newcommand{\bk}{{\bf k}}
\newcommand{\bQ}{{\bf Q}}
\newcommand{\q}{{\bf q}}
\newcommand{\s}{{\bf s}}
\newcommand{\bN}{{\bf \nabla}}
\newcommand{\bA}{{\bf A}}
\newcommand{\bE}{{\bf E}}
\newcommand{\bj}{{\bf j}}
\newcommand{\bJ}{{\bf J}}
\newcommand{\bs}{{\bf v}_s}
\newcommand{\bn}{{\bf v}_n}
\newcommand{\bv}{{\bf v}}
\newcommand{\la}{\langle}
\newcommand{\ra}{\rangle}
\newcommand{\dg}{\dagger}
\newcommand{\br}{{\bf{r}}}
\newcommand{\brp}{{\bf{r}^\prime}}
\newcommand{\bq}{{\bf{q}}}
\newcommand{\hx}{\hat{\bf x}}
\newcommand{\hy}{\hat{\bf y}}
\newcommand{\bS}{{\bf S}}
\newcommand{\cU}{{\cal U}}
\newcommand{\cD}{{\cal D}}
\newcommand{\bR}{{\bf R}}
\newcommand{\pll}{\parallel}
\newcommand{\sumr}{\sum_{\vr}}
\newcommand{\cP}{{\cal P}}
\newcommand{\cQ}{{\cal Q}}
\newcommand{\cS}{{\cal S}}
\newcommand{\ua}{\uparrow}
\newcommand{\da}{\downarrow}
\newcommand{\red}{\textcolor {red}}

\title{Nonequilibrium fluctuation theorems in the presence of a
time-reversal symmetry-breaking field and nonconservative forces}

\author{Punyabrata Pradhan}

\affiliation{ Physics Department, Technion - Israel Institute of
Technology, Haifa, Israel}

\begin{abstract}
\noindent We study nonequilibrium fluctuation theorems for
classical systems in the presence of a time-reversal
symmetry-breaking field and nonconservative forces in a stochastic
as well as a deterministic set up. We consider a system and a heat
bath, called the combined system, and show that the fluctuation
theorems are valid even when the heat bath goes out of equilibrium
during driving. The only requirement for the validity is that,
when the driving is switched off, the combined system relaxes to a
state having a uniform probability measure on a constant energy
surface, consistent with microcanonical ensemble of an isolated
system.

\typeout{polish abstract}
\end{abstract}


\maketitle

\section{Introduction}

Understanding thermodynamics of irreversible processes from time
reversible microscopic dynamics  has been a subject of great
interest since the time of foundation of statistical physics. A
fair amount of progress has been made since then, especially
through the recently discovered fluctuation theorems \cite{Evans,
review1, review2, review3, Evans_Searles, Gallavotti}. The
fluctuation theorems give a quantitative measure of
irreversibility in terms of asymmetries in the probability
distributions of various quantities in a driven system, e.g., heat
produced in a sheared fluid \cite{Evans, Evans_Searles,
Gallavotti}, heat exchanged between a hot and a cold body being in
contact \cite{JarzynskiPRL2004, SaitoPRL2007}, etc. These
nonequilibrium quantities, usually termed as `entropy production'
in an irreversible process \cite{Kurchan, Lebowitz, Sasa, Maes,
Seifert}, are on average non-negative and give an insight into the
2nd law of thermodynamics. The fluctuation theorem was originally
derived for deterministic thermostatted dynamics \cite{Gallavotti}
and later for stochastic one, such as Langevin dynamics
\cite{Kurchan} and Markovian jump processes \cite{Lebowitz}.

Closely connected to the fluctuation theorems, there are two
remarkable relations, called the Jarzynski equality
\cite{JarzynskiPRL1997, Park_Schulten, Dellago} and the Crooks
theorem \cite{CrooksPRE1999}, which involve fluctuation of work
done on a system driven arbitrarily far away from equilibrium by
varying an external parameter. Consider a system which is
initially at an equilibrium state ${\cal A}$ at temperature $T$
and coupled to an external parameter $\lambda(t)$. The system is
driven out of equilibrium by varying $\lambda(t)$ in a time
interval $0 \le t \le \tau$ where $\lambda(t)$ is constant outside
this time interval. In this process, called forward process for a
fixed protocol $\lambda(t)$, an amount of work $W$ is done on the
system. The system eventually relaxes to an equilibrium state
${\cal B}$ at same temperature $T$. In the reverse process, the
system is driven from the initial equilibrium state ${\cal B}$ for
a reverse protocol $\lambda(\tau-t)$ and eventually the system
relaxes to the equilibrium state ${\cal A}$. The Jarzynski
equality relates average of $\exp(-\beta W)$ performed over
nonequlibrium trajectories to equilibrium free energy difference
$\Delta F = F({\cal B})-F({\cal A})$ as \be \langle \exp(-\beta
W)\rangle = \exp(-\beta \Delta F) \ee where $F({\cal A})$ and
$F({\cal B})$ are equilibrium free energy of the system at state
${\cal A}$ and ${\cal B}$ respectively, and $\beta=1/k_B T$, $k_B$
the Boltzmann constant. The Crooks theorem relates the ratio of
the probabilities of work done $W$ for the forward process and
that for the reverse process, \be \frac{P_F(W)}{P_R(-W)} =
e^{\beta(W-\Delta F)} \ee where $P_F(W)$ and $P_R(W)$ are the
probability distributions of work for the forward and the reverse
processes, respectively.

Dissipative mechanism of a heat bath is crucial for understanding
irreversible phenomena and the fluctuation theorems \cite{Blythe},
and is modeled in various ways, such as, by employing
deterministic thermostatted dynamics \cite{Evans_Searles,
Gallavotti}, stochastic Langevin dynamics satisfying the
fluctuation-dissipation theorem \cite{Kurchan, Narayan, Mai_Dhar,
Speck_Seifert} or Markov dynamics satisfying detailed balance with
respect to canonical measure \cite{CrooksPRE1999, Crooks2000}.
However in these cases, the heat bath is not considered
explicitly, and is assumed to be always in equilibrium. In a
realistic scenario, heat generated by a driving force is
continuously dissipated to the heat bath, and consequently the
portion of the heat bath in the vicinity of the system goes away
from equilibrium during driving \cite{CohenJStatMech,
CohenMolPhys}.

It is therefore important how one employs a heat bath to take into
account the nonequilibrium effect of the bath. Recently there is a
prescription of modeling a driven system, possibly in contact with
a nonequilibrium heat bath, by applying Jaynes' principle of
entropy-maximization \cite{Jaynes} to nonequilibrium trajectories
with a macroscopic flux constraint \cite{RMLEvans_PRL,
RMLEvans_JPhysA, Simha}. However we follow a different path where
we explicitly consider a system and a heat bath combined, either
obeying microscopic Newtonian dynamics or obeying stochastic
dynamics with symmetries of the microscopic Newtonian dynamics
preserved. The work fluctuation relations have been studied along
this line before for classical Hamiltonian dynamics
\cite{Jarzynski_JStatPhys, Broeck} as well as stochastic dynamics
\cite{Pradhan}, but any time-reversal symmetry-breaking fields or
nonconservative forces have not been considered. Recently, the
fluctuation theorem involving particle-current has been studied in
a case of quantum mechanical transport of electrons across a
quantum dot in the presence of a time-independent magnetic field
\cite{Saito_Utsumi}.

In this paper, we generalize the fluctuation theorems for
classical systems in the presence of a time-reversal
symmetry-breaking field, such as an external magnetic field, and
nonconservative forces which cannot be derived from gradient of
scalar potentials. We consider a system and a heat bath, combined,
in a deterministic as well as a stochastic set up and we show that
the fluctuation theorems are valid in the presence of
time-reversal symmetry-breaking fields and nonconservative forces,
even when the heat bath goes out of equilibrium during the
driving. The validity only requires that (1) in the absence of
driving the system and the heat bath, combined, relax to a state
with a uniform probability measure on a constant energy surface,
and (2) there exists a time-reversal operation under which work
performed on the system is odd. Although we specifically consider
an external magnetic field in the paper, the results are also
applicable to other time-reversal symmetry-breaking fields, e.g.,
a Coriolis force present in a rotating system.

In a deterministic set up, we consider a system obeying Newtonian
dynamics and we prove the fluctuation theorems using the fact that
Liouville's theorem is valid even in the presence of an external
time-dependent magnetic field as well as other nonconservative
forces. We also extend our analysis to stochastic dynamics in the
presence of a time-reversal symmetry-breaking field in a
microcanonical set up. We consider an isolated system governed by
Markovian dynamics where there is violation of detailed balance
with respect to a uniform measure, i.e., forward and corresponding
reverse transition probabilities are not equal in general.
Reversing the time-reversal symmetry-breaking field results in
dynamics where all forward and corresponding reverse transition
probabilities are interchanged with each other. Although detailed
balance is violated, we prove the fluctuation theorems only
requiring that the steady state measure of an isolated system is
uniform on a constant energy surface.

We primarily focus on the work fluctuation relations, i.e., the
Jarzynski equality and the Crooks fluctuation theorem. The system
is driven by varying a control parameter of an external potential
or (and) by nonconservative forces. For systems obeying Newtonian
dynamics, the driving force may be due to a nonconservative
electric field induced by an external time varying magnetic field
in addition to other nonconservative forces. The work fluctuation
theorems have recently been studied for a few specific cases of a
single Brownian particle in the presence of both time-independent
\cite{Jayannavar, Roy} and time-dependent \cite{Saha_Jayannavar}
magnetic field. However we formulate the problem in a more general
setting, taking into account the system and the heat bath degrees
of freedom explicitly. Note that our analysis is applicable only
to the classical systems consisting of particles which do not have
any intrinsic magnetic moment.

Here is a brief outline of the paper. In section II, we study
systems obeying Newtonian dynamics in the presence of an external
magnetic field and some other nonconservative force fields. In
section III, we give a general proof of the fluctuation theorems
for stochastic systems in the absence of detailed balance in a
microcanonical set up, in section IV we then illustrate the ideas
using two simple stochastic models. In section V, we generalize
the results for other intensive thermodynamic variables, e.g.,
pressure and chemical potential which determine the initial and
final equilibrium states of the system in contact with a heat
bath.

\section{Newtonian dynamics}

First, we study the nonequilibrium fluctuation theorems in a
general deterministic framework for a system and a heat bath
combined, called the {\it combined system} (CS). We consider the
CS, which is governed by microscopic Newtonian dynamics, in the
presence of an external magnetic field $\vec{B}(\vec{r}, t)$ and a
nonconservative force field $\vec{f}(\vec{r}, t)$. Force fields in
general may be dependent both on position $\vec{r}$ and time $t$.
The nonconservative force $\vec{f}(\vec{r}, t)$ cannot be derived
from gradient of a scalar potential. In addition, there may be
conservative forces present in the CS which can be derivable from
gradient of scalar potentials. A microstate of the CS is denoted
by a variable ${\bf Y}$ which contains positions and velocities of
all particles, i.e., ${\bf Y} \equiv (\vec{r}_1, \vec{r}_2, \dots,
\vec{v}_1, \vec{v}_2, \dots) \equiv \{ \vec{r}_i, \vec{v}_i\}$
where $\vec{r}_i$ and $\vec{v}_i$ are position and velocity of
$i$-th particle in the CS respectively. Newton's equations of
motion for $i$-th particle can be written as \bea \dot{\vec{r}}_i
= \vec{v}_i \mbox{~,~~~~~~~~~~~~~~~~~~~~~~~~~~~~~~~~~~~~~~~~}
\label{Newton1}
\\
m_i \dot{\vec{v}}_i = - \vec{\nabla}_{\vec{r}_i} V(\vec{r}_i,
\lambda(t)) + q_i \vec{v}_i \times \vec{B}(\vec{r}_i, t) \nonumber
\\
- q_i \frac{\partial \vec{A}(\vec{r}_i, t)}{\partial t}  +
\vec{f}(\vec{r}_i, t) \label{Newton2} \eea where $m_i$ and $q_i$
are mass and charge of the $i$-th particle, $V(\vec{r}, \lambda(t)
)$ is total scalar potential at position $\vec{r}$ due to the
inter-particle interaction potentials as well as an external
potential with  a time-dependent control parameter $\lambda(t)$,
$\vec{\nabla}_{\vec{r}_i}$ the gradient operator with respect to
coordinate $\vec{r}_i$, $\vec{A}(\vec{r}, t)$ is the vector
potential at position $\vec{r}$ and time $t$ due to the external
magnetic field $\vec{B}(\vec{r}, t)$ which can be written as curl
of the vector potential, i.e., $\vec{B}(\vec{r}, t) = \nabla
\times \vec{A}(\vec{r}, t)$. The third term in the r.h.s. of Eq.
\ref{Newton2} is due to the time varying magnetic field
$B(\vec{r}, t)$ which induces a nonconservative electric field,
$-{\partial \vec{A}}/{\partial t}$. The induced electric field,
like $f(\vec{r},t)$, cannot be derived from gradient of a scalar
potential.

It is important to note that, when the nonconservative force field
$f(\vec{r}, t)$ is present, there is {\it no} Hamiltonian for the
CS and consequently Eqs. \ref{Newton1}, \ref{Newton2} {\it cannot}
be derived using a familiar Hamiltonian prescription of classical
mechanics \cite{Goldstein}. However the microscopic Newtonian
equations of motion are still invariant under time-reversal with
the direction of the magnetic field also reversed, i.e., as $t
\rightarrow -t$, $\vec{v}_i \rightarrow -\vec{v}_i$ and
$\vec{B}(\vec{r}) \rightarrow - \vec{B}(\vec{r})$ (equivalently
$\vec{A}(\vec{r}) \rightarrow - \vec{A}(\vec{r})$), Eqs.
\ref{Newton1} and \ref{Newton2} remain unchanged. Therefore, for
any trajectory ${\bf Y}(t) \equiv \{ \vec{r}_i(t), \vec{v}_i(t)\}$
with a fixed protocol $\{ \lambda(t), \vec{B}(\vec{r}, t),
\vec{f}(\vec{r}, t) \}$ in a time range $-{\cal T} \le t \le {\cal
T}$, there exists a reverse trajectory $\tilde{{\bf Y}}(t) \equiv
\{ \vec{r}_i(-t), -\vec{v}_i(-t)\}$ for the corresponding reverse
protocol $\{ \lambda(-t), -\vec{B}(\vec{r},-t), \vec{f}(\vec{r},
-t) \}$. Note that in the time-reversal operation mentioned above,
{\it direction} of the nonconservative forces, $f$ as well as
$-{\partial A}/{\partial t}$, are {\it unchanged}.

In the subsequent discussions, we consider a process in a time
interval $-{\cal T} \le t \le {\cal T}$ where ${\cal T}$ is very
large compared to any other time scales. We assume that the
magnetic field $\vec{B}(\vec{r}, t)$, the external parameter
$\lambda(t)$ and the nonconservative force field $\vec{f}(\vec{r},
t)$ couple only to the system. The field $\vec{B}(\vec{r}, t)$ (or
equivalently the vector potential $\vec{A}(\vec{r}, t)$),
$\lambda(t)$ and $\vec{f}(\vec{r}, t)$ are varied according to a
fixed protocol only in time interval $0 \le t \le \tau$ where
$\tau \ll {\cal T}$. Otherwise $\vec{B}$, $\lambda$ are kept
constant and $\vec{f}=0$ outside the interval $0 \le t \le \tau$.

Although there is no Hamiltonian in the presence of
nonconservative forces, energy function of the CS, in terms of
positions and velocities, can be defined as \be E(\{ \vec{r}_i,
\vec{v}_i\}) = \left( \sum_i \frac{1}{2} m_i \vec{v}_i^2 \right) +
V(\{ \vec{r}_{i}\}, \lambda(t)), \label{E_CS} \ee where the first
term is the total kinetic energy and the second term $V(\{
\vec{r}_{i}\}, \lambda(t))$ is the total potential energy,
containing both the interaction pair-potentials dependent on
relative position $|\vec{r}_i - \vec{r}_j|$ between any pair of
particles $i$, $j$ and an external potential with a control
parameter $\lambda$. The total energy of the CS, defined in terms
of positions and velocities, {\it does not depend} on the external
magnetic field. However a time varying magnetic field does change
the energy of the CS because of the work performed by an induced
nonconservative electric field ${\partial
\vec{A}(\vec{r}_i,t)}/{\partial t}$. This can be seen as
following: Using Eqs. \ref{Newton1}, \ref{Newton2}, the rate of
change of total energy $E$ of the CS can be written as \be
\frac{d}{dt} \left[ E(\{ \vec{r}_i, \vec{v}_i\}) \right] = \sum_i
\vec{v}_i.\vec{{\cal F}_i} + ({\partial V}/{\partial \lambda})
\dot{\lambda}, \ee where $\vec{{\cal F}_i} = [- q_i {\partial
\vec{A}(\vec{r}_i,t)}/{\partial t}+\vec{f}(\vec{r}_i,t)]$ is sum
of all the external nonconservative forces acting on $i$-th
particle of the CS. This implies that the rate of change of total
energy of the CS equals to the rate of work $W$ done by all the
external forces on the system, i.e., $(dE/dt) = (dW/dt)$. Total
work $W$ performed on the system can be calculated as $W =
\int_{-{\cal T}}^{\cal T} (dE/dt) dt$, or \be W = \int_0^{\tau}
\left( \frac{dE}{dt} \right) dt \ee since $(dE/dt)=0$ outside the
time interval $0\le t \le \tau$. The rate of change of energy is
clearly odd under time-reversal, i.e., $(dE/dt) \rightarrow
-(dE/dt)$ as $t \rightarrow -t$, $\vec{v}_i \rightarrow
-\vec{v}_i$ and $\vec{A} \rightarrow -\vec{A}$ because $(\partial
\vec{A}/\partial t)$ is even and $d\lambda/dt$ is odd under
time-reversal. In other words, total work performed equals to the
difference in total energy between the final and the initial point
of a trajectory and therefore total work is odd under
time-reversal, ${\cal W}_F[{\bf Y}(t)] = - {\cal W}_R[\tilde{\bf
Y}(t)]$.

Let us now consider time evolution of phase space density
$\rho({\bf Y})$ at a phase space point ${\bf Y} \equiv\{
\vec{r}_i, \vec{v}_i \}$. From the equation of continuity, one
obtains that the rate of change of phase space density $\rho({\bf
Y})$  equals to the divergence of local phase space current
density $\rho \dot{\bf Y}$, i.e., one gets the local conservation
equation \cite{Tolman} \be \frac{\partial \rho}{\partial t} +
\frac{\partial}{\partial {\bf Y}} (\rho \dot{\bf Y}) = 0 ,
\label{ConEq1} \ee where we have denoted the divergence of the
phase space current as $\partial (\rho \dot{\bf Y})/\partial {\bf
Y} = \sum_{i,\beta} [ \partial (\rho \dot{r}_{i,\beta})/\partial
{r_{i,\beta}} + \partial (\rho \dot{v}_{i,\beta})/\partial
{v_{i,\beta}} ]$ where $r_{i,\beta}$ and $v_{i,\beta}$ are
$\beta$-th Cartesian component ($\beta=1,2,3$ in three dimension)
of the position vector $\vec{r}_{i}$ and the velocity vector
$\vec{v}_{i}$ respectively. Taking derivative explicitly with
respect to the phase space point ${\bf Y}$, Eq. \ref{ConEq1} can
be rewritten as \be \frac{\partial \rho}{\partial t} + \dot{\bf
Y}.\left( \frac{\partial \rho} {\partial {\bf Y}} \right) + \rho
\frac{\partial \dot{\bf Y}}{\partial {\bf Y}} =0, \label{ConEq2}
\ee where $\dot{\bf Y}.({\partial \rho}/{\partial {\bf Y}}) =
\sum_{i,\beta} [ \dot{r}_{i,\beta} (\partial \rho/\partial
r_{i,\beta}) + \dot{v}_{i,\beta} (\partial \rho/\partial
v_{i,\beta}) ]$ and the phase space compression factor ${\partial
\dot{\bf Y}}/{\partial {\bf Y}} = \sum_{i,\beta} [ (\partial
\dot{r}_{i,\beta}/\partial r_{i,\beta}) + (\partial
\dot{v}_{i,\beta}/\partial v_{i,\beta}) ]$. Since the r.h.s. of
Eq. \ref{Newton1} is independent of $\vec{r}_i$, taking partial
derivative of Eq. \ref{Newton1} with respect to the position
coordinate, one gets $(\partial \dot{r}_{i,\beta}/\partial
r_{i,\beta})=0$. Note that the 2nd term in the r.h.s. of Eq.
\ref{Newton2} depends on $\vec{v}_i$ only through the cross
product with the external magnetic field vector $\vec{B}$ and
therefore partial derivative $\partial (\vec{v}_i \times
\vec{B})_{\beta}/\partial v_{i,\beta} = 0$ where $(\vec{v}_i
\times \vec{B})_{\beta}$ denotes $\beta$-th Cartesian component of
the vector $(\vec{v}_i \times \vec{B})$. Since all other terms in
the r.h.s. of Eq. \ref{Newton2} are independent of $\vec{v}_i$,
taking derivative of Eq. \ref{Newton2} with respect to the
velocity coordinates, one gets $(\partial
\dot{v}_{i,\beta}/\partial v_{i,\beta}) = 0$. This implies that
the phase space compression factor $\partial {\bf
\dot{Y}}/\partial {\bf Y}=0$, and therefore, from Eq.
\ref{ConEq2}, one arrives at Liouville's theorem, \be
\frac{d\rho}{dt} = \left[ \frac{\partial \rho}{\partial t} +
\dot{\bf Y}. \left( \frac{\partial \rho} {\partial {\bf Y}}
\right) \right] = 0. \ee The above equation is an important
statement which says that, even in the presence of a
time-dependent external magnetic field and other time-dependent
nonconservative forces, a set of phase space points flow like an
incompressible fluid under microscopic Newtonian time evolution
equations. Given that the phase space is incompressible, the CS at
$t=\pm{\cal T}$, with $\vec{f}=0=-({\partial \vec{A}}/{\partial
t})$ and $\lambda =constant$, can be considered to have a uniform
(microcanonical) measure on a constant energy surface, $E({\bf Y},
\lambda)=constant$.

One can  now prove the Crooks theorem by using Liouville's theorem
that the phase space is incompressible and the property that total
work performed on the CS is odd under simultaneous reversal of
time and the magnetic field. Let us denote the probability
distributions of work $W$, $P(W; \alpha(t)) \equiv P_F(W)$ and
$P(W; \tilde{\alpha}(t)) \equiv P_R(W)$, respectively for a
forward protocol $\alpha(t) \equiv \{ \lambda(t), \vec{f}(\vec{r},
t), \vec{B}(t)\}$ and corresponding reverse protocol
$\tilde{\alpha}(t) \equiv \{ \lambda(-t), \vec{f}(\vec{r}, -t),
-\vec{B}(-t)\}$. Now following the arguments along the line of
Ref. \cite{Broeck, Pradhan}, we consider a set $A$ of initial
phase space points at time $t=-{\cal T}$ which evolve from a
constant energy surface with energy $E$ to a set of points $A'$ of
the final phase space points of a constant energy surface with
energy $E+W$ at time $t={\cal T}$ for driving under the forward
protocol $\alpha(t)$. Total work performed on the system in this
process is $W$ and the probability distribution of work
$P_F(W)=\omega(A)/\Omega(E)$ where $\omega(A)$ is the phase space
volume of the set $A$ and $\Omega(E)$ is the phase space volume of
the constant energy surface with total energy $E$. Now for any
trajectory with the forward protocol $\alpha(t)$, there exists a
unique time reversed trajectory with the reverse protocol
$\tilde{\alpha}(t)$, and work performed along a time reversed
trajectory, initially starting from one of the set of phase space
points $A'_{R}$ obtained by velocity-reversal of the set $A'$, is
negative of the work performed for the corresponding forward
trajectory. Therefore, for the reverse trajectories, the phase
space transforms from the energy surface with energy $E+W$ to an
energy surface with energy $E$. Then, the probability distribution
$P_R(-W)$ can be written as $P_R(-W)=\omega(A'_{R})/\Omega(E+W)$.
Now using Liouville's theorem that phase space is incompressible,
we have $\omega(A)=\omega(A')$, and then using
$\omega(A')=\omega(A'_{R})$ that phase space volume does not
change under reversal of velocities, one obtains the ratio of the
probabilities of work $W$ and $-W$ as
$P_F(W)/P_R(-W)=\Omega(E+W)/\Omega(E)$. The ratio can also be
written as \be \frac{P_F(W)}{P_R(-W)} = \frac{P_{st}({\bf
Y}({-\cal T}), \lambda(0))}{P_{st}({\bf Y}({\cal T}),
\lambda(\tau))} \label {Crooks_Newtonian} \ee where $P_{st}({\bf
Y}, \lambda)=1/\Omega(E({\bf Y}, \lambda))$ is the initial or the
final equilibrium probability distribution of the CS. Note that
$P_{st}({\bf Y}, \lambda)$ is {\it independent of the external
magnetic field} $\vec{B}$ as the total energy $E$ given in Eq.
\ref{E_CS}, and therefore $\Omega(E)$, does not depend on
$\vec{B}$.

At this point one can separate the system from the heat bath by
defining entropy and temperature of the CS which has a uniform
probability measure on a constant energy surface. The probability
$P_{st} ({\bf Y},\lambda)$ of a microstate of the CS, at $t=\pm
\cal{T}$, is inverse of phase space volume $\Omega(E)$ of a
constant energy surface with energy $E$, i.e.,
$P_{st}=1/\Omega=\exp(-S/k_B)$ where $S$ is defined as entropy. We
set the Boltzmann constant $k_B=1$ afterwards. Partitioning the CS
into two parts, the system and the heat bath with energies
$\epsilon$ and $(E-\epsilon)$ respectively, one can write $P_{st}
({\bf Y},\lambda)= \left[ \int e^{S_{B}(E-\epsilon)+ S(\epsilon,
\lambda)} d\epsilon \right]^{-1}$ where $S_B(E-\epsilon)$ and
$S(\epsilon, \lambda)$ are entropy of the heat bath and the system
respectively. We have here assumed the interaction energy between
the system and the bath to be much smaller than energy of either
the system or the bath. Now introducing inverse temperature
$\beta$ of the heat bath, $\beta={\partial S_B(E)}/{\partial E}$
and expanding $S_B(E-\epsilon)$ in leading order of $\epsilon/E$,
$S_B(E-\epsilon) = S_B(E) - \beta \epsilon + {\cal O}(\epsilon/E)
$ in the limit $\epsilon \ll E$, one gets $P_{st}(E, \lambda) =
e^{-S_B(E)} e^{\beta F(\lambda)}$ where the Helmholtz free energy
of the system $F(\lambda)=-(1/\beta) \ln \left[ \int e^{-\beta
\epsilon} e^{S(\epsilon, \lambda)} d\epsilon \right]$ with
$e^{S(\epsilon, \lambda)}$ density of states of the system with
energy $\epsilon$.

From conservation of energy, we have $E({\cal T}) = [E({-\cal T})
+ W]$ where $E({-\cal T})=E({\bf Y}({-\cal T}), \lambda_0)$,
$E({\cal T})=E({\bf Y}({\cal T}), \lambda_{\tau})$ and $W$ is
total work performed for the forward protocol. Writing
probabilities of the initial  and final microstates respectively
as $P_{st}({\bf Y}({-\cal T}), \lambda_0) = e^{-S_B(E({-\cal T}))}
e^{\beta F(\lambda(0))}$ and $P_{st}({\bf Y}({\cal T}),
\lambda({\tau})) = e^{-S_B(E({-\cal T}))} e^{-\beta W} e^{\beta
F(\lambda({\tau}))}$, one gets the ratio of probabilities of the
final and initial equilibrium microstates of the CS as \be
\frac{P_{st}({\bf Y}({-\cal T}), \lambda(0))}{P_{st}({\bf Y}({\cal
T}), \lambda(\tau))} = e^{\beta (W - \Delta F)} \label{ratio_prob}
\ee where $\beta$ is inverse equilibrium temperature of the heat
bath. Note that, writing ${P_{st}({\bf Y}({-\cal T}),
\lambda(0))}/{P_{st}({\bf Y}({\cal T}), \lambda(\tau))} =
\exp(\Delta S_{CS})$ in the l.h.s of Eq. \ref{ratio_prob}, one
obtains the thermodynamic relation $T \Delta S_{CS} = W - \Delta
F$ where $\Delta S_{CS}$ is change in total entropy $S_{CS}$ of
the CS, $\Delta F$ is change in free energy of {\it only} the
system and temperature $T=1/\beta$ \cite{dT_bath}. Now
substituting the above ratio of the probabilities into Eq.
\ref{Crooks_Newtonian}, one obtains the Crooks theorem in the
presence of a time-dependent external magnetic field and a
nonconservative force, \be \frac{{P(W; \lambda(t),
\vec{B}(t))}}{{P(-W; \lambda(-t), {-\vec{B}}(-t))}} =
e^{\beta(W-\Delta F)}. \ee The Jarzynski equality $\langle
\exp(-\beta  W) \rangle = \exp(-\beta \Delta F)$ follows by
integrating the Crooks theorem \cite{CrooksPRE1999}.

\begin{figure}
\begin{center}
\leavevmode
\includegraphics[width=8cm,angle=0]{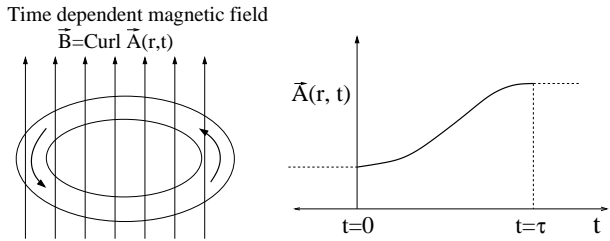}
\caption{Electric current flows in a ring due to a time-dependent
magnetic field $\vec{B}(\vec{r}, t)=\nabla \times \vec{A}(\vec{r},
t)$, perpendicular to the ring, where the vector potential
$\vec{A}(\vec{r}, t)$ at position $\vec{r}$ varies with time $t$.}
\label{At}
\end{center}
\end{figure}

The Crooks theorem has a simpler form when $\lambda$ is kept
constant (implying $\Delta F=0$), $\vec{f}=0$ throughout and only
the magnetic field $\vec{B}(t)$ varies in a time-symmetric cycle
where $\vec{B}(t) = \vec{B}(\tau-t)$ with initial and final values
of $\vec{B}=0$. Consider an electrical circuit which is symmetric
with respect to $\vec{B}(t)$, e.g., see Fig. \ref{At} where a ring
is placed in an uniform time-dependent magnetic field in the
direction perpendicular to the ring. The time varying magnetic
field induces an oscillating electric field and an electric
current in the circuit. For any finite number of such cycles, the
induced electric field $-\partial \vec{A}/\partial t$ performs
work $W$ on the system and thus generates heat in the circuit. In
this case, due to the geometric symmetry, the probability
distribution of work is same for $\vec{B}(t)$ and $-\vec{B}(t)$
and only depends on the magnitude of $\vec{B}$, i.e.,
$P_F(W)=P_R(W) \equiv P(W; |\vec{B}|)$. Since $\vec{B}$ varies in
time-symmetric cycle, one finally arrives at the Crooks theorem
${P(W; |\vec{B}|)}/{P(-W; |\vec{B}|)} = \exp(\beta W)$ which gives
an estimate of irreversibility of the heat produced in an
alternating electric current-carrying circuit.

The fluctuation theorems can be similarly extended to the cases
where there are a Coriolis force $2 m (\vec{\omega} \times
\vec{v})$ and a centrifugal force $m \vec{\omega} \times
(\vec{\omega} \times \vec{r})$ acting on a particle of mass $m$
\cite{Goldstein}, $\vec{\omega}$ being angular velocity of the
rotating system, in addition to an external magnetic field. The
fluctuation theorems are still valid provided that one reverses
the direction of the angular velocity $\vec{\omega}$ as well as
the magnetic field $\vec{B}$. This is because even if one adds the
centrifugal and Coriolis forces in Eq. \ref{Newton2}, the phase
space is still incompressible, i.e., $\partial {\bf
\dot{Y}}/\partial {\bf Y}=0$.

\section{Stochastic dynamics}

In this section, we consider a system and a heat bath combined
(CS), in a general stochastic framework. Stochasticity may arise
due to incomplete knowledge of some of the degrees of freedom in
the original deterministic system \cite{deGroot}. Due to
incomplete knowledge of the degrees of freedom, a system is
described by some coarse-grained variables and is governed by a
stochastic dynamics. We consider a Markovian dynamics of the CS,
specified by transition probability $w({\bf Y}'|{\bf Y}) d{\bf
Y}'$, from a configuration ${\bf Y}$ at time $t$ to any
configuration in the volume element $d{\bf Y}'$ around ${\bf Y}'$
at time $t+\Delta t$, where the degrees of freedom of the CS are
denoted as ${\bf Y}(t)$ at time $t$. Transitions are allowed on a
constant energy surface of the CS. In subsequent discussion, we
consider a class of models where, in the absence of driving, a
uniform (microcanonical) measure is realized on a constant energy
surface of the isolated CS.

Transition probabilities  are chosen so that they obey symmetries
and conservation laws of underlying microscopic dynamics. The
degrees of freedom ${\bf Y}(t) \equiv \{{\bf Y}_+, {\bf Y}_-\}$
may be identified as two sets of stochastic variables, ${\bf Y}_+$
(e.g., position) and ${\bf Y}_-$ (e.g., velocity), and there
exists a $\bar{\bf Y} \equiv \{{\bf Y}_+, -{\bf Y}_-\}$ for any
given ${\bf Y}$. In the presence of a time-reversal
symmetry-breaking field, such as an external magnetic field
$\vec{B}$, we impose a condition on the transition probabilities
as given below, \be w({\bf Y}'|{\bf Y}; \vec{B}) = w(\bar{{\bf
Y}}|\bar{{\bf Y}}'; -\vec{B}). \label{MR} \ee The above condition
can be taken as definition of a magnetic field in a stochastic set
up where reversing the magnetic field results in interchanging
forward and corresponding reverse transition probabilities with
each other. Indeed, under suitable assumptions, the transition
probabilities chosen above can be derived for a closed isolated
classical system governed by a microscopic Newtonian dynamics
\cite{vanKampen, deGroot}. Note that Eq. \ref{MR} equates
transition probabilities of two {\it different} systems, one with
a magnetic field $\vec{B}$ and the other with a magnetic field
$-\vec{B}$. Time-reversal of a trajectory ${\bf Y}(t)$, in a
symmetric time range $-{\cal T} \le t \le \cal{T}$, is defined as
$\tilde{{\bf Y}}(t) = \bar{\bf Y}(-t) \equiv \{ {\bf Y}_+(-t),
-{\bf Y}_-(-t)\}$ when $t \rightarrow -t$. The variables ${\bf Y}$
and $\bar{\bf Y}$ transform to each other under time-reversal
where time-reversal operation is ensured by the condition in Eq.
\ref{MR}.

In the absence of a magnetic field, Eq. \ref{MR} (with
$\vec{B}=0$) implies extended detailed balance condition $w({\bf
Y}'|{\bf Y}) = w(\bar{{\bf Y}}|\bar{{\bf Y}}')$. When ${\bf Y}
\equiv \{\bf{Y}_+\}$ contains only position-like variables, Eq.
\ref{MR} becomes $w({{\bf Y}_+}|{{\bf Y}'}_+; \vec{B})= w({{\bf
Y}'}_+|{{\bf Y}_+}; -\vec{B})$ which, for $\vec{B}=0$, implies the
condition of detailed balance $w({{\bf Y}_+}|{{\bf Y}'}_+)=
w({{\bf Y}'}_+|{{\bf Y}_+})$ \cite{vanKampen}.

Note that, under the condition of Eq. \ref{MR}, an isolated CS in
general {\it does not satisfy} detailed balance. To stress the
violation of detailed balance, later in section III.A, we would
specifically consider a case where the reverse transition is not
allowed and any reverse transition probability corresponding to a
forward one is set to be zero.

However choice of the transition probabilities cannot be arbitrary
and one has to put some constraints so that (a) all the states are
connected to each other ensuring the Markov process is ergodic,
and (b) steady state configurations are all equally probable. In
this paper, we consider a class of stochastic models which satisfy
constraints (a) and (b). For a network of discrete states in a
configuration space, a sufficient condition for such a class,
which we call {\it loopwise balance} condition, can easily be
formulated (see appendix for details). Even if the CS relaxes to a
state having a uniform measure, the state would be a
nonequilibrium steady state due to the violation of detailed
balance. It is important to note that, provided there exists a
unique and uniform steady state measure $\rho({\bf Y})=constant$
for a Markov process with a magnetic field $\vec{B}$, the Markov
process with the reverse magnetic field $-\vec{B}$ is well
defined, i.e., the transition probabilities are still normalized
$\int d{\bf Y}' w({\bf Y}'|{\bf Y}; -\vec{B})=1$, and the same
steady state measure $\rho({\bf Y})=constant$ is guaranteed for
the Markov process with $-\vec{B}$. In other words, $\int d{\bf Y}
w({\bf Y}'|{\bf Y}; \vec{B}) \rho({\bf Y}) = \rho({\bf Y}')=
constant \Rightarrow \int d{\bf \bar{Y}}' w(\bar{{\bf
Y}}|\bar{{\bf Y}}'; -\vec{B}) \rho(\bar{{\bf Y}}')= \rho(\bar{{\bf
Y}}) = constant$, i.e., {\it uniform steady state measure is
invariant under reversal of the magnetic field}. This is because
the normalization condition $\int d{\bf Y}' w({\bf Y}'|{\bf Y};
\vec{B}) = 1$ implies the steady state condition $\int d{\bf
\bar{Y}}' w(\bar{{\bf Y}}|\bar{{\bf Y}}'; -\vec{B})=1$, which can
be shown by using Eq. \ref{MR} and the transformation ${\bf Y}
\rightarrow \bar{{\bf Y}}$. This is discussed and illustrated in
the appendix.

We stress that the assumption of Markovian dynamics of a system
and a heat bath, combined, is weaker than that of Markovian
dynamics of only the system. Even if the combined system obeys
Markovian dynamics, dynamics of the system, in lower dimensional
configuration space, is {\it non-Markovian}, in contrast to the
system considered in Ref. \cite{Crooks2000}.

The total energy of the CS is denoted as $E({\bf Y}, \lambda)$
where $\lambda$ is an external parameter coupled only to the
system. When the CS is not driven, $E({\bf Y}, \lambda)$ is
conserved. Importantly, total energy $E$ depends explicitly only
on ${\bf Y}$ and $\lambda$, not on the magnetic field $\vec{B}$
\cite{vanVleck} and it is an even function of ${\bf Y}_-$ so that
$E({\bf Y}, \lambda)=E(\bar{\bf Y}, \lambda)$. For simplicity we
assume time $t$ changes in discrete step of $\Delta t$ and we
consider a Markov chain in a time range $-{\cal T} \le t \le {\cal
T}$ where ${\cal T}$ is very large. The parameter $\lambda$ is
changed  from $\lambda=\lambda_0$ to $\lambda=\lambda_{\tau}$
according to a deterministic protocol in a finite time interval $0
\le t \le \tau$ where $\tau \ll {\cal T}$ and otherwise kept
constant. We call it a forward protocol. A reverse protocol is
defined as $\{\tilde{\lambda}_t\} \equiv \{\lambda_{-t}\}$.

An amount of work $\delta W_t$ at time step $t$ may be performed
on the system in two ways. One may usually change the external
parameter from $\lambda_t$ to $\lambda_t + \delta \lambda_t$,
keeping ${\bf Y}$ fixed, and the work performed is $\delta W_t =
E({\bf Y}_t, \lambda_t+\delta \lambda_t) - E({\bf Y}_t,
\lambda_t)$. Now we introduce here the second way of performing
work on the system. One may also change the degrees of freedom of
the CS, at a time step $t$, deterministically from ${\bf Y}_t$ to
${\bf Y}'_t=S^{\Delta t}({\bf Y}_t)$, keeping $\lambda$ fixed,
where $S^{t}$ is a time-reversal symmetric evolution operator,
i.e., if ${\bf Y} \rightarrow {\bf Y}'$ under influence of a
nonconservative force, $\bar{{\bf Y}}' \rightarrow \bar{{\bf Y}}$
under influence of the same force. For example, $S^{t}$ may simply
be the Newtonian time evolution operator. We will illustrate this
by using a simple model in section IV.B. Work performed in this
case is calculated as $\delta W_t = E({\bf Y}_t+\delta {\bf Y}_t,
\lambda_t) - E({\bf Y}_t, \lambda_t)$ which is the work performed
by a nonconservative force when the evolution operator $S^{t}$
contains such a force. The total work $W$ performed on the system
is written as $W= \sum_t \delta W_t$.

A trajectory is denoted by $\{{\bf Y}_t, \lambda_t, \vec{B} \}$
where ${\bf Y}_t$, $\lambda_t$ are respective values of ${\bf Y}$,
$\lambda$ at time $t$ and $\vec{B}$ is the external magnetic
field. Given a trajectory $\{{\bf Y}_t, \lambda_t, \vec{B}\}$,
there is a unique reverse trajectory $\{\tilde{{\bf Y}}_t,
\tilde{\lambda}_t, - \vec{B}\}$ with reversed magnetic field
$-\vec{B}$ and reverse protocol $\{\tilde{\lambda}_t\}$. Note that
the trajectory $\{\tilde{{\bf Y}}_t, \tilde{\lambda}_t,
\vec{B}\}$, without reversing $\vec{B}$, may not even be
realizable if some of the reverse transition probabilities are
zero. From Eq. \ref{MR}, the probabilities of a trajectory from a
given initial configuration with the magnetic field $\vec{B}$ and
that of the corresponding reverse trajectory with $-\vec{B}$ are
equal, \be {\cal P}[\{{\bf Y}_t, \lambda_t, \vec{B}\}] = {\cal
P}[\{\tilde{{\bf Y}}_t, \tilde{\lambda}_t, -\vec{B}\}]
\label{MR_trajectory} \ee where ${\cal P}[.]$ denotes respective
probability of a trajectory. We call the above equation as the
microscopic reversibility (MR) condition hereafter. As a special
case, when $\vec{B}=0$, the above condition can be written simply
as \be {\cal P}[\{{\bf Y}_t, \lambda_t\}] = {\cal P}[\{\tilde{{\bf
Y}}_t, \tilde{\lambda}_t\}]. \label{MR_trajectory1} \ee

We define ${\cal W}_F[\{{\bf Y}_t, \lambda_t, \vec{B}\}]$ as work
performed along a trajectory $\{{\bf Y}_t, \lambda_t, \vec{B}\}$
where ${\cal W}_F = [E({\bf Y}_{\cal T}, \lambda_{\tau}) - E({\bf
Y}_{- \cal T}, \lambda_0)]$, the difference in total energy of the
final and initial point of the trajectory. For a forward protocol
$\{\lambda_t\}$, we define the probability distribution of work
$W$, as $P(W; \{\lambda_t \}, \vec{B}) \equiv P_F(W)$ which can be
written as \bea P_F(W) = \sum_{\{{\bf Y}_{t}\}} P_{st}({\bf
Y}_{-{\cal T}}, \lambda_0, \vec{B}_0) {\cal P}[\{{\bf Y}_t,
\lambda_t, \vec{B}\}] \nonumber \\
\times \delta({\cal W}_F - W), \label{Def_forward} \eea where
$P_{st}({\bf Y}_{-{\cal T}}, \lambda_0, \vec{B})$ is the initial
steady state distribution at time $t=-{\cal T}$, and ${\cal
P}[\{{\bf Y}_t, \lambda_t, \vec{B}\}]$ is the probability of the
trajectory $\{{\bf Y}_t, \lambda_t, \vec{B}\}$. For the reverse
protocol $\{ \tilde{\lambda}_t \}$ with reversed magnetic field
$-{\vec{B}}$, the probability distribution $P(W; \{\lambda_{-t}
\}, -\vec{B}) \equiv P_R(W)$ of work $W$ can be written as \bea
P_R(W) = \sum_{\{\tilde{{\bf Y}}_t \}} P_{st}(\bar{\bf Y}_{\cal
T}, \lambda_{\tau}, -\vec{B}_{\tau}) {\cal P}[\{ \tilde{{\bf Y}},
\tilde{\lambda}, -{\vec{B}} \}] \nonumber \\
\times \delta({\cal W}_R - W), \label{Def_reverse} \eea where work
performed along the trajectory $\{ \tilde{{\bf Y}}_t,
\tilde{\lambda}_t, -{\vec{B}} \}$ is ${\cal W}_R = [ E(\bar{\bf
Y}_{-\cal T}, \lambda_0) - E(\bar{\bf Y}_{\cal T}, \lambda_{\tau})
]$.

Throughout the paper, we use two symmetry relations as following.
\\
(1) ${\cal W}_F[\{{\bf Y}_t, \lambda_t, \vec{B}\}] = - {\cal
W}_R[\{\tilde{{\bf Y}}_t, \tilde{\lambda}_t, -{\vec{B}}
\}]$, i.e., the work performed is odd under simultaneous reversal
of time and the magnetic field.
\\
(2) $P_{st}({\bf Y}, \lambda, \vec{B}) = P_{st}({\bf Y}, \lambda)
= P_{st}(\bar{\bf Y}, \lambda)$, i.e., the steady state
distribution is independent of the magnetic field and invariant
when velocities are reversed.
\\
To show the symmetry relation 1, one should note that work done
along a trajectory is, by definition, the difference in total
energy of the CS at the final and the initial point of the
trajectory, implying ${\cal W}_F = [E({\bf Y}_{\cal T},
\lambda_{\tau}) - E({\bf Y}_{- \cal T}, \lambda_0)]$ and ${\cal
W}_R = [ E(\bar{\bf Y}_{-\cal T}, \lambda_0) - E(\bar{\bf Y}_{\cal
T}, \lambda_{\tau}) ]$. Now using $E({\bf Y}, \lambda) =
E(\bar{\bf Y}, \lambda)$, i.e., energy is invariant when
velocities are reversed, one obtains the symmetry relation 1. The
symmetry relation 2 holds because the steady state distribution of
the CS is uniform on a constant energy surface where energy of the
CS is independent of the magnetic field and the uniform steady
state distribution does not change for the reversed velocities.
Independence of the total energy on the magnetic field has already
been manifested in Eq. \ref{E_CS} where energy of a deterministic
system has been expressed in terms of positions and velocities
\cite{vanVleck}.

Using microscopic reversibility condition of Eq.
\ref{MR_trajectory} and the symmetry relation 2, changing
summation indices $\{{\bf Y}_{t}\} \rightarrow \{\tilde{{\bf
Y}}_t\}$, and then using the symmetry relation 1, Eq.
\ref{Def_forward} can be rewritten as \bea P_F(W) =
\sum_{\{\tilde{{\bf Y}}_t\}} \left( \frac{P_{st}({\bf Y}_{-{\cal
T}}, \lambda_0)}{P_{st}({\bf Y}_{\cal T}, \lambda_{\tau})} \right)
P_{st}(\bar{\bf Y}_{\cal T},
\lambda_{\tau}) \nonumber \\
\times {\cal P}[\{ \tilde{{\bf Y}}_t, \tilde{\lambda}_t, -\vec{B}
\}] \delta({\cal W}_R + W). \label{Crooks1} \eea Now defining
entropy and temperature of the CS, as done before in the case of
Newtonian dynamics in section II, one can write the probabilities
of the initial and final microstates respectively as $P_{st}({\bf
Y}_{-\cal T}, \lambda_0) = e^{-S_B(E_{-\cal T})} e^{\beta
F(\lambda_0)}$ and $P_{st}({\bf Y}_{\cal T}, \lambda_{\tau}) =
e^{-S_B(E_{-\cal T})} e^{-\beta {\cal W}_F} e^{\beta
F(\lambda_{\tau})}$ where $S_B$ entropy of the heat bath and
$F(\lambda)$ the Helmholtz free energy of the system with the
external parameter $\lambda$. So the ratio of the probabilities
can be written as \be \frac{P_{st}({\bf Y}_{-{\cal T}},
\lambda_0)}{P_{st}({\bf Y}_{\cal T}, \lambda_{\tau})} = e^{\beta
({\cal W}_F - \Delta F)}, \label{ratio} \ee where $\Delta F =
F(\lambda_{\tau}) - F(\lambda_0)$ the difference in the Helmholtz
free energy. Substituting the above ratio of probabilities into
Eq. \ref{Crooks1}, one arrives at the Crooks theorem in the
presence of an external magnetic field, \be \frac{P(W;
\{\lambda_t\}, \vec{B})}{P(-W; \{\tilde{\lambda}_t\}, -{\vec{B}})}
= e^{\beta (W - \Delta F)} \label{Crooks2} \ee where the
probability distributions of work in general depend on the
magnetic field $\vec{B}$ as the transition probabilities depend on
$\vec{B}$. The Jarzynski equality, $\langle \exp(-\beta W) \rangle
= \exp(-\beta \Delta F)$, is derived straightforwardly by
integrating the Crooks theorem \cite{CrooksPRE1999}. Note that Eq.
\ref{Crooks2} relates the probability distributions of work for
two systems with different microscopic dynamics, i.e., one system
with a magnetic field $\vec{B}$ and the other with a magnetic
field $-\vec{B}$. Importantly, unlike the Crooks theorem, the
Jarzynski equality is written without any reference to the
magnetic field and so the Jarzynski equality is a statement
regarding a system with a particular dynamics.

The Crooks theorem takes an interesting form, if geometry of a
system is symmetric with respect to the magnetic field $\vec{B}$.
Given this symmetry, the work probability distributions do not
depend on the direction of $\vec{B}$, but depend only on the
magnitude $|\vec{B}|$: $P(W; \{{\lambda_t}\}, \vec{B}) = P(W;
\{\lambda_t\}, -\vec{B}) \equiv P(W; \{\lambda_t\}, |\vec{B}|)$
(similarly for the work probability distribution with reverse
protocol $\{\tilde{\lambda}_t\}$). This implies that, in this
case, the Crooks theorem holds even when the magnetic field is
same for the forward and the reverse protocol. Replacing the index
$-\vec{B}$ by $\vec{B}$ in Eq. \ref{Crooks2}, one can now write
the Crooks theorem as  ${P(W; \{\lambda_t\}, \vec{B})}/{P(-W;
\{\tilde{\lambda}_t\},\vec{B})} = \exp[\beta (W - \Delta F)]$.
Note that in this case one does not have to reverse the direction
of the magnetic field in the reverse protocol, and therefore the
Crooks theorem expresses symmetries in the probability
distributions of work for a system with same dynamics for the
forward and the reverse protocol. This type of symmetry would be
illustrated in an example given in section IV.A.

\section{Stochastic Dynamics: Illustration}

In this section, we illustrate the ideas developed in the previous
section by constructing two simple stochastic models. First we
consider the effect of an external magnetic field where, to ensure
violation of detailed balance, we specifically choose reverse
transition probability to be zero for any nonzero forward
transition probability. Second we consider a nonconservative force
in a stochastic set up. Although the two models considered in this
section are just toy models for a system and a heat bath, they
nevertheless demonstrate the dissipative and equilibrating
mechanism of a heat bath, and subsequently show the validity of
the Crooks theorem even when the heat bath goes out of equilibrium
during driving.

\subsection{Time-reversal symmetry-breaking field}

\begin{figure}
\begin{center}
\leavevmode
\includegraphics[width=7cm,angle=0]{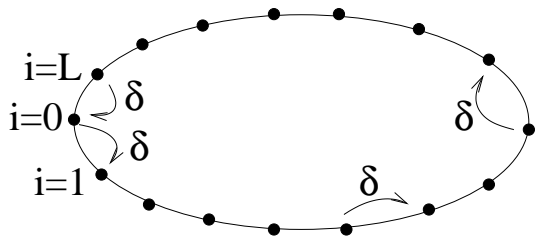}
\caption{Schematic diagram of the system and the heat bath. The system
is the site $i=0$, and rest of the sites, $1 \le i \le L$, constitute
the heat bath.}
\label{CS}
\end{center}
\end{figure}

We take a one dimensional ring of $L+1$ sites where site $i=0$ is
considered as the system and all other sites, $1 \le i \le L$, are
considered to be the heat bath (see Fig. \ref{CS}). At any site
$i$ there is an energy variable $e_i \ge 0$. The energy at site
$i=0$ is given by $e_0= \lambda x$ where the external parameter
$\lambda$ couples only to the system via an internal degree of
freedom $x >0$. A configuration of the CS is thus specified by
${\bf Y} \equiv \{x, e_1, \dots , e_L\}$. The dynamics is the
following: a site $i$ is chosen randomly and a fixed amount of
energy $\delta$ ($ \ll 1$) is transferred only in one direction
(say, anti-clockwise) to the nearest neighbor site, i.e., \be e_i
\rightarrow e_i - \delta \mbox{~} ; \mbox{~} e_{i+1} \rightarrow
e_{i+1} + \delta. \ee The total energy $E=\sum_{i=0}^L e_i$ is
conserved in this process. Whenever energy $e_0$ at $i=0$  is
changed, the variable $x$ is updated accordingly: $e_0 \rightarrow
e_0' \Rightarrow x \rightarrow x'=e_0'/\lambda$. For $e_i <
\delta$, the energy transfer is not allowed.

There is a mean energy current in anti-clockwise direction which
may be considered to be due to an externally applied field in this
direction (analogous to $\vec{B}$). The reverse field corresponds
to the dynamics where energy is transferred in clockwise direction
(analogous to $-\vec{B}$). Since there are no velocity-like
variables, we have ${\bf Y} = {\bf \bar{Y}}$ and time-reversal is
simply defined as ${\bf Y}(t) \rightarrow {\bf Y}(-t)$ as $t
\rightarrow -t$ in a symmetric time interval $-{\cal T} \le t \le
{\cal T}$. Note that, in this case, reverse transition probability
is zero for any nonzero forward transition probability because
energy is transferred only in one direction (anti-clockwise),
i.e., $w({\bf Y}|{\bf Y}'; \vec{B})=0$ for any nonzero $w({\bf
Y}'|{\bf Y}; \vec{B}) \ne 0$. Therefore a time reversed trajectory
is possible only for the dynamics where energy is transferred in
the reverse direction (i.e., clockwise). Clearly, the model
satisfies the microscopic reversibility $w({\bf Y}'|{\bf Y};
\vec{B}) = w({\bf Y}|{\bf Y}'; -\vec{B})$ as given in Eq. \ref{MR}
(with ${\bf Y} = {\bf \bar{Y}}$), and also satisfies the symmetry
relations 1 and 2.

When $\lambda$ is kept constant, total energy $E$ is conserved and
the dynamics is a totally asymmetric zero range process
\cite{tazrp} on a ring with a constant hopping rate where number
of particles at a site is $e_i/\delta$. With total number of
particles fixed in the process, steady state configurations are
all equally probable. This can be understood by mapping the zero
range process to a totally asymmetric simple exclusion process
\cite{tazrp} where all possible states are equally probable in the
steady state. In the limit of large $L$, probability distribution
of energy at any site $i$ is given by the Boltzmann distribution,
$P(e_{i}) = \beta e^{-\beta e_i}$ where $\beta=[\sum_{i=0}^{L}
e_i/(L+1)]^{-1}$ is inverse temperature of the CS. The partition
function of the system, for a fixed value of $\lambda$, can be
calculated as ${\cal Z} (\lambda) = \int_0^{\infty} e^{- \beta
\lambda x} dx = (\beta \lambda)^{-1}$ and the free energy is given
by $F(\lambda) = -\beta^{-1} \ln {\cal Z}$.

\begin{figure}
\begin{center}
\leavevmode
\includegraphics[width=9.0cm,angle=0]{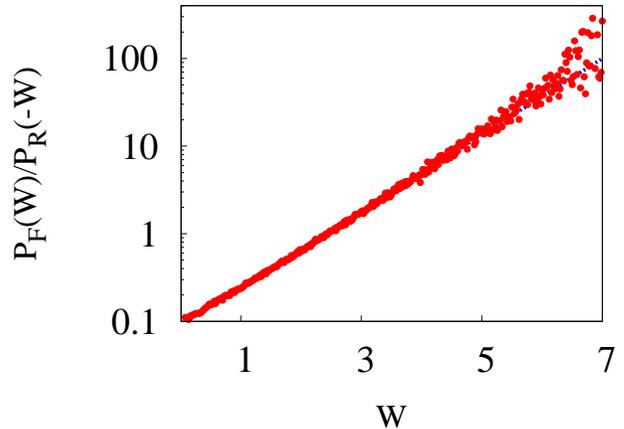}
\caption{The ratio $P_F(W)/P_R(-W)$ is plotted versus work $W$ in
semi-Log scale and fitted with $\exp[\beta (W-\Delta F)]$ for
$L=100$, $\beta=1.0$, $\lambda_0=1.0$, $\lambda_{\tau}=11.0$ where
$\tau=100$ and $\Delta F = (1/\beta) \ln
(\lambda_{\tau}/\lambda_0)$.} \label{Crooks}
\end{center}
\end{figure}

The system is driven by changing the external parameter, in
discrete step of $\delta \lambda_t$ at $t$-th time step, from an
initial value $\lambda_0$ to a final value $\lambda_{\tau}$ in
time interval $0 \le t \le \tau$. For each increment $\delta
\lambda_t$, an amount of energy $\delta W_t$ is added to the
system ($i=0$) where $\delta W_t = (\partial e_0/\partial \lambda)
\delta \lambda_t= x.\delta \lambda_t$ is defined as work performed
at $t$-th time step. Total work performed is $W=\sum_t \delta
W_t$. We set a unit of time such that all sites are updated with
rate one per unit Monte Carlo time. For the reverse protocol, the
external parameter is varied as
$\tilde{\lambda}(t)=\lambda(\tau-t)$ in time interval $0 \le t \le
\tau$, from $\lambda_{\tau}$ to $\lambda_0$. Note that energy is
always transferred in anti-clockwise direction both for the
forward and the reverse protocol. The probability distributions of
work $W$ for the forward and the reverse protocol are denoted as
$P(W; \lambda) \equiv P_F(W)$ and $P(W; \tilde{\lambda}) \equiv
P_R(W)$ respectively. Due to symmetriy of the ring geometry, the
work distributions do not depend on the direction of the energy
transfer. We verify numerically that the Crooks theorem is indeed
satisfied, i.e., $P_F(W)/P_R(-W) = \exp[\beta(W-\Delta F)]$ with
$\Delta F = F(\lambda_{\tau}) - F(\lambda_0)$. In Fig.
\ref{Crooks}, we plot $P_F(W)/P_R(-W)$ as a function of $W$ where
$\lambda_0=1.0$, $\lambda_{\tau}=11.0$, $\tau=100$, $\beta=1.0$,
$L=100$. The parameter $\lambda$ is increased in a specific way:
first $\lambda$ is increased in $5$ equal discrete steps upto
$t=5$, then held constant upto $t=95$, and again increased in $5$
equal discrete steps upto $t=100$. The parameter $\lambda$ is
varied in this particular way to ensure that the energy
fluctuations travel around the ring and can perturb the system at
site $i=0$ within the measurement time $\tau=100$. In Fig.
\ref{Crooks}, the ratio $P_F(W)/P_R(-W)$ fits well with
$\exp[\beta(W-\Delta F)]$ where $\Delta F = (1/\beta)\ln
(\lambda_{\tau}/\lambda_0)$ is the theoretical of the difference
in free energy.

\subsection{Nonconservative force}

When a nonconservative force is present in the CS, e.g., a system
of particles in a ring in contact with a heat bath and with a
force acting in anti-clockwise direction as in Fig. \ref{At}, the
force field cannot be derived from the gradient of a scalar
potential and therefore cannot be absorbed in the expression of
the total energy of the CS (e.g., see Eq. \ref{E_CS}). In this
case, unlike changing an external parameter $\lambda$, the system
is driven by changing ${\bf Y} \rightarrow {\bf Y} + \delta {\bf
Y}$ as discussed in section III. To illustrate this, we consider a
CS which consists of $L+1$ lattice sites in one dimension. The
site $i=0$ has energy $e_0=p^2$, with an internal variable $p$,
and any other site $i$ has energy $e_i \ge 0$. The site $i=0$ is
considered to be the system and the rest is the heat bath. A
configuration of the CS is specified by ${\bf Y} \equiv \{p, e_1,
\dots, e_L\}$ where, for any given ${\bf Y}$, there is a $\bar{\bf
Y} \equiv \{-p, e_1, \dots, e_L\}$. The dynamics is chosen as
follows. For $1 \le i \le L$ we choose a site at random and
exchange energy between sites $i$ and $i+1$ randomly, \be e_i
\rightarrow q (e_i+e_{i+1}); \mbox{~~} e_{i+1} \rightarrow
(1-q)(e_i+e_{i+1}), \label{NC1} \ee where $q \in [0,1]$ is a
uniform random number. The total energy $E=\sum_{i=0}^L$ is
constant in this process. We update the site $i=0$ slightly
differently where we consider that the site $i=0$ can interchange
energy only with site $i=1$. Say, energy of the two sites, before
update, are $e_0=p^2$ and $e_1$ respectively. We generate a random
number $\xi$ uniformly distributed in the range
$[-p_{max},p_{max}]$ where $p_{max}=\sqrt{e_0+e_1}$. We then
update the internal variable $p$ and energy of the site $i=0,1$ as
given below, \be p \rightarrow \xi; \mbox{~~} e_0 \rightarrow
\xi^2; \mbox{~~} e_1 \rightarrow (e_0+e_1-\xi^2). \label{NC2} \ee
The update rule ensures that detailed balance is satisfied with
respect to a uniform measure on a constant energy surface of the
CS. Consequently, while the CS is not driven, the site $i=0$ has
the Boltzmann probability distribution $P(p)=\exp(-\beta
p^2)/{\cal Z}$ where $\beta=[\sum_{i=0}^{L} e_i/(L+1)]^{-1}$ is
inverse temperature of the CS and ${\cal
Z}=\int_{-\infty}^{\infty} dp \exp(-\beta p^2)$ is the partition
function.

The system is driven by changing the internal variable $p$ as
follows: $p \rightarrow p + \delta$ where $\delta >0$ is a
constant (choice of the sign of $\delta$ is arbitrary). Now two
following steps performed repeatedly: Step.1 - random sequential
update of $L$ bonds of the CS using Eq. \ref{NC1}, \ref{NC2} and
Step.2 - update of the site $i=0$ by changing the internal
variable from $p$ to $p + \delta$. The second step may be thought
of, as if the internal variable $p$ is like momentum of a particle
and it is updated due to effect of an external constant
nonconservative force $f$ which changes $p$ by a fixed amount
$\delta = f.dt$ in a small time interval $dt$. Note that, under
the driving, the transition $p \rightarrow p + \delta$ (also $-(p
+ \delta) \rightarrow -p$) is allowed, but the transition $p +
\delta \rightarrow p$ is not allowed. In other words, internal
variable $p$ only changes in one direction, i.e., either increases
(for $\delta>0$) or decreases (for $\delta<0$), under the driving.
Work $\delta W_t$ done on the system is the change in energy of
the system (i.e., site $i=0$) where $\delta W_t = (p+\delta)^2
-p^2=\delta(2p+\delta)$ and total work $W=\sum_t\delta W_t$.

The dynamics considered above is similar to the Langevin dynamics
of a Brownian particle in a thermal environment where an external
nonconservative force is acting on the particle. One should note
that, given a trajectory $[p(t),\{e_i(t)\}]$, one can define
time-reversal operation in two ways, i.e., as $t \rightarrow -t$,
$[p(t),\{e_i(t)\}] \rightarrow [-p(-t),\{e_i(-t)\}]$ or
$[p(t),\{e_i(t)\}] \rightarrow [p(-t),\{e_i(-t)\}]$. But only the
first way of time-reversal is relevant here because, given a
trajectory $[p(t),\{e_i(t)\}]$, the trajectory
$[p(-t),\{e_i(-t)\}]$ is not realizable as $p$ only increases
under the driving. However the microscopic reversibility condition
in Eq. \ref{MR_trajectory1} is satisfied as the transition
probabilities have an additional symmetry,
$w(p',\{{e_i}'\}|p,\{e_i\})=w(-p,\{e_i\}|-p',\{{e_i}'\})$, i.e.,
$w({\bf Y}'|{\bf Y}) = w(\bar{\bf Y}|\bar{\bf Y}')$.

Following the general proof given in section III, one can see that
the Crooks theorem is satisfied, $P(W)/P(-W)=\exp(\beta W)$, where
the free energy change $\Delta F=0$. Since the forward and reverse
protocol of driving is same in the above example, we have used
$P_F(W) \equiv P_R(W) \equiv P(W)$ in the Crooks theorem. For a
time-dependent external nonconservative force, the increment
$\delta_t$ of the internal variable $p$ at a time step $t$ will be
$\delta_t = f_t.dt$, where $f_t$ is the force at time step $t$. In
this case, the reverse protocol should be $\{f_{-t}\}$ for a given
forward protocol $\{f_t\}$, and one should distinguish between the
work probability distributions $P_F(W)$ and $P_R(W)$. Then the
Crooks theorem can be written in the more general form as
$P_F(W)/P_R(-W)=\exp(\beta W)$.

\section{Generalization}

The fluctuation theorems can be generalized to the cases where a
system is in contact with a heat bath with pressure $P$ and (or)
chemical potential $\mu$. Let us consider the combined system with
total energy $E$, volume $V$ and number of particles $N$ which are
globally conserved. Energy, volume and number of particles
$\epsilon$, $v$ and $n$ of the system fluctuate due to interaction
with the heat bath. Pressure $P$ and chemical potential $\mu$ can
be defined, similar to temperature, as given below, \bea
\beta P = \frac{\partial S_B(E, V, N)}{\partial V}, \\
\beta \mu = \frac{\partial S_B(E, V, N)}{\partial N}. \eea Now
using the expansion of the heat bath entropy $S_B(E-\epsilon, V-v,
N-n) = S_B(E,V, N) - \beta \epsilon - \beta P v -\beta \mu n$ in
the limit of $\epsilon \ll E$, $v \ll V$ and $n \ll N$, one can
rewrite the ratio of the probabilities of microstates at $t=\pm
{\cal T}$, as given in Eq. \ref{ratio}, as \be \frac{P_{st}({\bf
Y}_{-{\cal T}}, \lambda_0)}{P_{st}({\bf Y}_{\cal T},
\lambda_{\tau})} = e^{\beta ({\cal W}_F[\{{\bf Y}_{t},
\lambda_{t}, \vec{B}_{t}\}] - \Delta {\cal G})} \label{ratio2} \ee
where ${\cal G}(\beta, P, \mu, \lambda)$ the grand potential of
the system in equilibrium with a heat bath of inverse temperature
$\beta$, pressure $P$ and chemical potential $\mu$, $\lambda$ an
external parameter and $\Delta {\cal G} = {\cal G}(\lambda_{\tau})
- {\cal G}(\lambda_0)$ with the grand potential ${\cal G}$ defined
as \be {\cal G} (\lambda)=-\frac{1}{\beta} \ln \left[ \int
d\epsilon \int dv \int dn e^{-\beta (\epsilon + Pv + \mu n)}
e^{S(\epsilon, v, n, \lambda)} \right] \label{F2} \ee where
$S(\epsilon, v, n, \lambda)$ is entropy of the system. Then the
Crooks theorem can be written as given below, \be \frac{P(W;
\lambda(t), \vec{B}(t))}{ P(-W; \lambda(-t)_, -\vec{B}(-t))} =
e^{\beta (W - \Delta {\cal G})} \ee which is obtained by replacing
the Helmholtz free energy $F$ in Eq. \ref{Crooks2} by the grand
potential ${\cal G}$.

\section{Summary}

In this paper, we have studied the fluctuation theorems for a
classical system in contact with a heat bath in the presence of a
time-reversal symmetry-breaking field and nonconservative forces,
in a deterministic as well as a stochastic set up. We have shown
that the fluctuation theorems are valid under the condition that,
in the absence of any driving, the system and the heat bath,
combined, relax to a state having a uniform probability measure on
a constant energy surface. The fluctuation theorems have been
proved in a very general setting by using the time-reversal
symmetry and the conservation laws, and accordingly defining the
intensive thermodynamic variables like temperature, pressure,
chemical potential obtained from a microcanonical ensemble. In the
deterministic case of Newtonian dynamics, we have first shown that
Liouville's theorem holds even in the presence of a time-dependent
external magnetic field and other time-dependent nonconservative
forces and then, using Liouville's theorem, we have proved the
Crooks Theorem and the Jarzynski equality in the presence of such
forces. In the stochastic case, where the combined system obeys
Markovian dynamics, the work fluctuation theorems have been shown
to be valid even when the reverse transition probabilities are not
equal to the corresponding forward transition probabilities, thus
violating detailed balance condition.

\section{Acknowledgments}

\begin{acknowledgments}
The author thanks J. Robert Dorfman, Dov Levine and Yariv Kafri for many
useful discussions and acknowledges a fellowship of the Israel Council for
Higher Education.
\end{acknowledgments}

\section{Appendix}

For an equilibrium system, detailed balance with respect to a
uniform (microcanonical) probability measure is a sufficient
condition for all states to be equally probable in the final
equilibrium state. Here we formulate a sufficient condition for
having equally probable steady states for a nonequilibrium system
with finite number of states.

\begin{figure}
\begin{center}
\leavevmode
\includegraphics[width=8.5cm,angle=0]{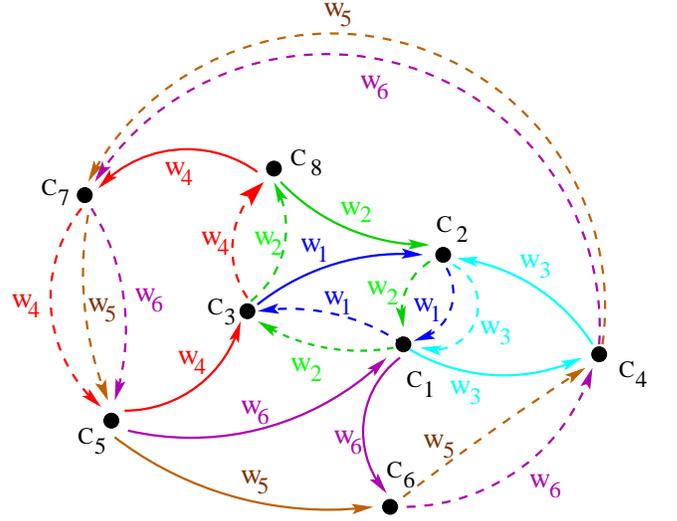}
\caption{(Color online) Schematic diagram of a network in a
configuration space: Configurations are denoted as nodes $C_1$,
$C_2$, $C_3$, $\dots$ $C_8$. Nodes are connected by various closed
loops, each of which is assigned a transition rate. Transition
rates assigned to the dotted arrows should be added to get the
corresponding total transition rate.} \label{network1}
\end{center}
\end{figure}

\begin{figure}
\begin{center}
\leavevmode
\includegraphics[width=8.5cm,angle=0]{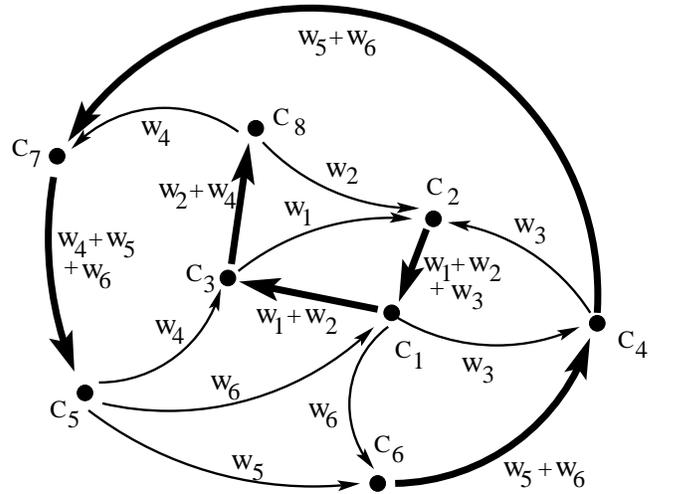}
\caption{Network with the magnetic field $\vec{B}$: Transition
rates assigned to the dotted arrows in Fig. \ref{network1} are
added to get the actual transition rate (denoted by the thick
arrows in this figure). Transition rates in general depend on the
magnetic field.} \label{network2}
\end{center}
\end{figure}

Although we now specifically consider the case where all reverse
transition rates to be zero for corresponding nonzero forward
transition rates (similar to the example considered in section
IV.A), the following discussion can be straightforwardly
generalized to cases where a forward and corresponding reverse
transition rate both may be nonzero. Also we only consider here
the case where there is no velocity-like variables, however the
generalization to such cases is straightforward. In Fig.
\ref{network1}, a network in a configuration space is shown
schematically. A configuration $C$ is denoted by a node in the
graph. Nodes are connected by drawing closed loops, where each
loop is assigned a transition rate, e.g., see Fig. \ref{network1}
where loops are assigned transition rates $w_1$, $w_2$, $w_3$,
etc. If two configurations are connected by more than one loop,
each assigned with different transition rates, the total
transition rate from one configuration to another is given by sum
of the transition rates. For example, in Fig \ref{network1}, the
total transition rate from $C_2$ to $C_1$ is $w(C_1|C_2) =
(w_1+w_2+w_3)$. Similarly, $w(C_3|C_1) = (w_1+w_2)$, $w(C_5|C_7) =
(w_4+w_5+w_6)$, etc. The resulting network is shown in Fig.
\ref{network2}. We call this way of assigning a transition rate
(or transition probability) to a closed loop of configurations in
a graph as {\it loopwise balance}. Only constraint for drawing
such loops is that all nodes must be connected to each other along
some path so that the system is ergodic. Apart from this, loops
are otherwise drawn arbitrarily. Note that since the Markov
process is ergodic, it has a unique steady state solution. There
are several ways to connect nodes satisfying the constraint of
having uniform steady state measure and, since the Markov process
is ergodic, all configurations always have equal steady state
probabilities. To see this, consider the Master equation for the
Markov process defined on a network in Fig. \ref{network1}, \bea
\frac{dP(C_1)}{dt} = -(w_1+w_2+w_3+w_6)P(C_1) \nonumber
\\
+ (w_1+w_2+w_3) P(C_2) +
w_6 P(C_5),
\nonumber
\\
\frac{dP(C_2)}{dt} = -(w_1+w_2+w_3)P(C_2)
\nonumber
\\
+ w_1 P(C_3) + w_2 P(C_8) + w_3 P(C_4),
\nonumber
\\
\dots
\nonumber
\\
\dots
\nonumber
\\
\frac{dP(C_7)}{dt} = -(w_4+w_5+w_6) P(C_7)
\nonumber
\\
+ (w_5+w_6) P(C_4) +
w_4 P(C_8),
\nonumber
\\
\frac{dP(C_8)}{dt} = -(w_2+w_4) P(C_8)
\nonumber
\\
+ (w_2+w_4) P(C_3). \eea From above set of equations it is clear
that all steady states have equal probabilities, i.e.,
$P(C_1)=P(C_2)=P(C_3)= \dots = P(C_8)=constant$ is the steady
state solution of the Master equation. Since the network is
ergodic, the steady state is also unique. Therefore {\it loopwise
balance} is a sufficient condition for having a uniform steady
state measure in an ergodic Markov process with finite number of
states.

\begin{figure}
\begin{center}
\leavevmode
\includegraphics[width=8.5cm,angle=0]{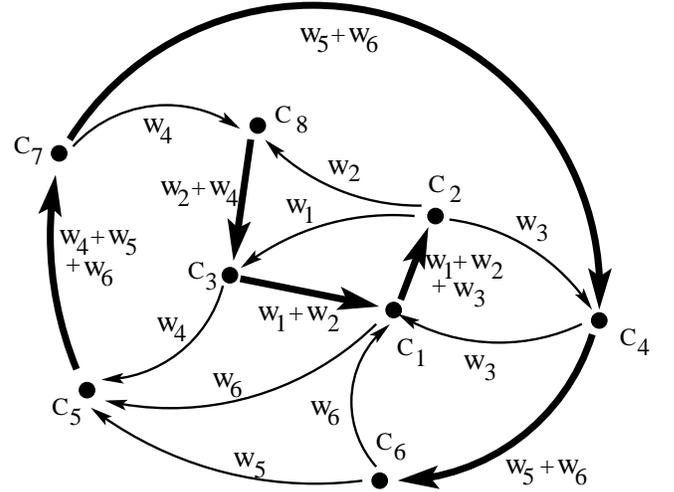}
\caption{Network with the reverse magnetic field $-\vec{B}$:
Reversal of the direction of the magnetic field $\vec{B}$
corresponds to reversal of all transition rates, i.e.,
interchanging forward and reverse transition rate. The network for
the reverse magnetic field $-\vec{B}$ results from the network of
Fig. \ref{network2} by reversing all the arrows.} \label{network3}
\end{center}
\end{figure}

If the Markov process, as defined on the network in Fig.
\ref{network2}, is considered to be in the presence of a magnetic
field $\vec{B}$, then the Markov process with the reverse magnetic
field $-\vec{B}$ is defined on the same network by assigning
transition rates from one node to another in the reverse
direction, i.e., just by reversing the arrows on a network as done
in Fig. \ref{network3}. Note that the transition rates assigned to
loops in general depend on the magnetic field. However the steady
state distribution remains uniform and thus independent of the
magnetic field, which is the symmetry relation 2 considered in
section III. Note that, although the Master equation changes under
reversal of the magnetic field, the steady state solution is still
unchanged, i.e., all steady states are still equally probable.


\begin{thebibliography}{99}
\bibitem{dT_bath} Since some amount of work done on the system gets
dissipated to the heat bath, there is a slight change $dT$ in
temperature of the bath
(although $dT$ is infinitesimal due to large size of a bath). The
net free energy change of the CS equals $\Delta F_{CS} = \Delta F_B +
\Delta F$ where $\Delta F_B = - S_{B} dT$ is a finite change
in free energy of the bath and $S_B$ is entropy
of the bath \cite{Narayan}.
\end{thebibliography}
\end{document}